\documentclass[10pt,twocolumn,preprintnumbers,superscriptaddress,aps,pra]{revtex4-2}

\usepackage[utf8]{inputenc}
\usepackage{amsmath, amssymb}
\usepackage{graphicx}

\usepackage{amsthm}
\usepackage{booktabs}
\usepackage{enumitem}
\usepackage{tabularx}
\usepackage{adjustbox}
\usepackage{cancel}

\theoremstyle{plain}
\newtheorem{theorem}{Theorem}[section]       
\newtheorem{corollary}[theorem]{Corollary}   
\newtheorem{lemma}[theorem]{Lemma}

\theoremstyle{definition}
\newtheorem{definition}[theorem]{Definition} 

\theoremstyle{remark}

\newcommand{\mlab}{MIT Media Lab}
\newcommand\MIT{Massachusetts Institute of Technology}
\newcommand{\bu}{Boston College}
\newcommand{\tu}{Tokyo University}

\begin{document}

\title{Propagational Proxy Voting}

\author{Yasushi Sakai}
\email{yasushi@mit.edu}
\affiliation{\mlab}
\affiliation{\MIT}

\author{Parfait Atchade-Adelomou}
\email{parfait@mit.edu}
\affiliation{\mlab}
\affiliation{\MIT}

\author{Ryan Jiang}
\email{jiangba@bc.edu}
\affiliation{\bu}

\author{Luis Alonso}
\email{alonsolp@mit.edu}
\affiliation{\mlab}
\affiliation{\MIT}

\author{Kent Larson}
\email{kll@mit.edu}
\affiliation{\mlab}
\affiliation{\MIT}

\author{Ken Suzuki}
\email{ken@sacral.c.u-tokyo.ac.jp}
\affiliation{\tu}

\date{April 2025}

\begin{abstract}
This paper proposes a voting process in which voters allocate fractional votes to their expected utility in different domains: over proposals, other participants, and sets containing proposals and participants. This approach allows for a more nuanced expression of preferences by calculating the result and relevance within each node. We modeled this by creating a voting matrix that reflects their preference. We use absorbing Markov chains to gain the consensus, and also calculate the influence within the participating nodes. We illustrate this method in action through an experiment with 69 students using a budget allocation topic.


\end{abstract}
\maketitle

\section{Introduction}

Collective decision-making processes are complex due to their multidimensional nature and different levels of knowledge. One form of collective decision-making, voting, is often coarsened into a small number of simplified options or left to highly chained delegation, causing a lack of control.
In this paper, we model a voting process in which participants allocate fractional votes based on their expected utility across different domains: over proposals, other participants, and sets containing the two. This approach allows for a more nuanced expression of preferences while allowing diversity in the level of participation. We model this in three steps by first modeling Liquid Democracy using absorbing Markov chains. Then we extend this, which was first introduced by Suzuki\cite{suzuki2013nameteki}. Finally, we add intermediaries to increase the flexibility of delegation. We test this model by applying it to an example case of budget allocation using Participatory Budgeting.

This work is structured as follows.
In Section~\ref{sec:problem}, we formally define the problem and discuss the theoretical foundations of our approach, highlighting the limitations of traditional voting methods and opportunities for innovation.
Section~\ref{sec:prior} reviews the latest participatory decision-making frameworks and utility-based voting models, situating our contribution within this broader context. 
Section~\ref{sec:modeling} details the proposed framework, describing its mathematical foundations and key components, including the integration of preference alignment, delegation dynamics, and influence metrics. 
Section~\ref{sec:utility} derives a utility based on the constant elasticity of substitution function.
Section~\ref{sec:validation} validates the framework through a case study using Cambridge Participatory Budgeting 2023 as a topic, demonstrating how PPV could reveal the complexity and guide participatory processes. 
Section~\ref{sec:discussion} explores the implications of the results, including potential applications, performance scalability, and limitations of the model. 
Finally, Section~\ref{sec:conclusion} summarizes the contributions.

\section{Problem Definition}\label{sec:problem} 

When a group makes a decision or evaluates, processes often become complex due to their multidimensional nature and diversity of participants. For example, Weyl et al. \cite{weyl2024plurality} highlight the complexity of land ownership in the San Francisco Bay Area, particularly for high-value properties. Land value is often shaped by contributions from a range of entities: agents, groups, jurisdictions (city, state, nation-state, and industry), and stakeholders, each operating at different scales, complicating the notion of singular ownership. These challenges are further complicated by hierarchical structures, both among participants and within the available options, which reflect interdependencies and nonlinear trade-offs. As Hayek pointed out \cite{hayek2013use}, the decentralized nature of currency systems serves as an information compression mechanism, distilling diverse inputs into a single price. However, in domains external to currency systems, the use of knowledge is especially costly and the resulting representations are often coarse and incomplete. In addition, the diverse levels of knowledge among the participants will also contribute to the complexity.

Voting, which is a form of collective decision-making, simplifies the above complexity by often reducing preferences to a single domain, either direct democracy or representative(indirect) democracy. 

Most participatory budgeting processes often require condensing a large number of proposals into shorter\footnote{The roll is called Budget Delegates or PB Delegates, in which your task is to create a shortlist.}, more manageable selections. In practical use cases, the above approach risks excluding valuable options, while compromising inclusivity and representativeness. To address these limitations requires a voting framework that respects multidimensional preferences, while supporting hierarchical decision-making, and ensures adaptability to diverse participant knowledge levels. This work proposes a decision-making framework that mirrors the principles of Liquid Democracy (\textit{LD}), yet extends the idea using each participant's expected utility scores. The model balances direct participation with flexible delegation, offering a dynamic approach to hierarchical decision-making. 

\section{Prior Work}\label{sec:prior}

Earlier contributions to the study of delegation mechanisms include Miller’s proposal \cite{miller1969program} of systems that allow voters to choose between direct and representative voting, as well as proxy voting models \cite{alger2006voting} that allow individuals to delegate their vote to another voter. 
Yet perhaps the closest reference is systems that combine direct and indirect participation, which have been explored to increase flexibility in democratic decision-making. Liquid democracy (LD), as described by Paulin \cite{paulin2014through}, enables individuals to vote directly or delegate their voting power to representatives. This model has been studied for its potential to balance efficiency and accountability in governance structures. Further research has expanded on the theoretical and practical implications of LD \cite{green2015direct, blum2016liquid, brill2018pairwise, caragiannis2019contribution, paulin2020overview, pierson2015outnumbered}, with several pilot implementations in various settings \cite{kling2015voting, hainisch2016civicracy, behrens2014principles, hardt2015google, daliparthi2023visdm}. Recent explorations include multiple delegations through ranked choice to avoid vote cycles \cite{utke2023anonymous}.

A key distinction of the Propagation Proxy Voting (PPV) model concerning LD is its treatment of vote splitting and the probabilistic nature of delegation. Whereas the LD literature generally formalizes the transitivity of votes through deterministic algorithms \cite{golz2018fluid, kahng2021liquid}, PPV allows a single vote to be distributed probabilistically between multiple targets and domains. This introduces additional flexibility in the capture of individual preferences and complex representation structures.

To our knowledge, this work builds on the model developed by Suzuki \cite{suzuki2013nameteki} written in Japanese, which applies absorbing Markov chains \cite{chung1967markov, grinstead1997introduction} to derive a social welfare function from individual preference matrices. Suzuki’s formulation provides a mathematical basis for modeling delegation as a propagation process. This paper extends this model by introducing intermediaries, defined as sets of policies or groups of delegates, as alternative delegation targets. This generalization enables the representation of hierarchical structures commonly found in electoral systems and policy-making processes, which are not explicitly addressed in Suzuki's work.

The introduction of intermediaries aligns this approach with methods designed for complex decision-making environments. A comparable example is the Analytic Hierarchy Process (AHP) \cite{saaty2008decision}, which decomposes decisions into hierarchical levels and uses pairwise comparisons to evaluate alternatives. Although AHP has been adapted for group decision-making \cite{saaty1989group}, it requires precise knowledge assessments at each level, which can be difficult to obtain in practice. Incorporating delegation mechanisms, as proposed in PPV, offers an alternative means of managing varying levels of expertise within hierarchical decision structures.

\section{The model}\label{sec:modeling}

\subsection{Step one: Liquid Democracy}

For our purposes, we first focus on modeling Liquid Democracy, as described by \cite{blum2016liquid} defining with absorbing Markov chains\cite{chung1967markov}. LD allows participants to vote directly or delegate their votes to peers, offering flexibility.

\begin{definition}[Liquid Democracy Model]
Let $N = \{d_1, d_2, \ldots, d_n\}$ be the set of voters, where $|N| = n$, and let $P = \{p_1, p_2, \ldots, p_k\}$ be the set of policies, where $|P| = k$.  We define the set of \emph{alternatives} as the union of voters and policies $A = N \cup P, \quad |A| = m$, where \( m = n + k \), assuming \( N \cap P = \emptyset \).
Let $\mathcal{V}$ be a voting matrix in this liquid democracy system, where each element $\mathcal{V}_{ij}$ indicates the vote cast on a target $i$ (either a participant or a policy) of the voter $j$. The matrix $\mathcal{V}$ adheres to the following constraints:

    \begin{align}
        \sum_{i=1}^{m} \mathcal{V}_{ij} &= 1 \quad \mathrm{for}\ \mathrm{all}\  1\leq j \leq A, \label{eq:LD_vote_sum_constraint} \\
        \mathcal{V}_{ii} &= 0 \Rightarrow i \in N, \label{eq:LD_no_vote_self_if_delegate} \\
        \mathcal{V}_{ii} &= 1 \Rightarrow i \in P, \label{eq:LD_vote_self_if_policy} \\
        \mathcal{V}_{ij} &\in \{0, 1\} \quad \mathrm{for}\ \mathrm{all}\ i\ \mathrm{and}\ j \label{eq:LD_binary_constraint}
    \end{align}
    This model illustrates the delegation of two domains: policies and peer participants, yet the whole vote needs to be allocated to a single target.The constraint \eqref{eq:LD_vote_sum_constraint} makes the matrix stochastic and guarantees that each voter's total voting power is conserved and fully allocated within the voting system.
\eqref{eq:LD_binary_constraint} ensures that the delegation of votes is binary: a vote is delegated or not. 
    \label{def:LD}
\end{definition}

\subsection{Step two: Propagational Proxy Voting}\label{sec:ppv}

\begin{definition}[Propagational Proxy Voting (PPV)]\label{def:PPV}
In the context of the Propagational Proxy Voting (PPV) model \cite{suzuki2013nameteki}, we define a voting matrix $\mathcal{V}$ that encapsulates the fundamental structure of a transposed absorbing Markov chain \cite{grinstead1997introduction} as follows:

\begin{equation}
\mathcal{V}_{ij} \in \mathbb{R}, \quad 0 \leq \mathcal{V}_{ij} \leq 1, \label{eq:PPV_fraction}
\end{equation}

Here, the components of the matrix \( \mathcal{V}' \) are defined as follows:

\begin{equation}
    \mathcal{V}' = \left(
    \begin{array}{c|c}
    \mathcal{V}_{d \leftarrow d} & \mathbf{0} \\
    \hline
    \mathcal{V}_{p \leftarrow d} & I 
    \end{array}
    \right).
    \label{eq:Voting_mat_def}
\end{equation}

\begin{itemize}
    \item \( \mathcal{V}_{d\leftarrow d} \) is the sub-matrix representing the transition probabilities \textit{between voters}. This block is a $n$ square matrix where $n$ is the total number of participants. Each element \( \mathcal{V}_{i,j} \) (\( 1 \leq i, j \leq n \)) represents the expected utility that voter \( j \) delegates to voter \( i \). Notably, the diagonal elements (\( \mathcal{V}_{i,i} \)) are constrained to be zero, ensuring that voters do not cast votes for themselves, as required by the model.

    \item \( \mathcal{V}_{p\leftarrow d} \) is the sub-matrix capturing the transition probabilities \textit{from participants(delegates) to policies}. This block has dimensions \( k \times n \), where \( k \) represents the total number of policies. Each element \( \mathcal{V}_{p_i,j} \) (\( n+1 \leq i \leq m \), \( 1 \leq j \leq n \)) indicates the distribution of fractional votes from participants \( j \) to a policy \( i \). 
    
    \item \( \mathbf{0} \) is a null matrix of dimensions \( n \times k \), enforcing the restriction that policies cannot directly cast votes toward participants. This reflects a key constraint of the model, whereby the flow of influence is directed from voters to voters or from voters to policies, but never directly from policies to voters.

    \item \( I \) is the identity matrix of size $k$, representing the \textit{absorbing states} in the Markov chain. Each diagonal element \( I_{i,i} \) (\( n+1 \leq i \leq m \)) is equal to 1. This ensures that once the voting influence reaches a policy, it remains there, effectively halting further propagation. This adheres to \eqref{eq:LD_vote_self_if_policy} in the LD definition.
\end{itemize}

This structure reflects the hierarchical nature of the voting process, with voters and policies occupying distinct roles. The four sections: \( \mathcal{V}_{d\leftarrow d} \), \( \mathcal{V}_{p\leftarrow d} \), \( \mathbf{0} \), and \( I \) ensures that the model adheres to the theoretical constraints while enabling a dynamic flow of influence within these parts.

In comparison to the \textit{LD} model, the strict binary constraint \eqref{eq:LD_binary_constraint} has been relaxed to the fractional constraint \eqref{eq:PPV_fraction}. 

This perspective highlights how votes propagate from transient to absorbing states, culminating in the final vote distribution.

Explicitly, with the information from \eqref{eq:Voting_mat_def}, the voting matrix $\mathcal{V}$ can be represented as:

\begin{equation}\label{eq:voting_matrix_expanded}
\mathcal{V} = 
\left(
\begin{array}{ccc|ccc}
\mathcal{V}_{1,1} & \cdots & \mathcal{V}_{1,n}  & 0 & \cdots & 0 \\
\vdots & \ddots & \vdots  & \vdots & \ddots & \vdots \\
\mathcal{V}_{n,1} & \cdots & \mathcal{V}_{n,n}  & 0 & \cdots & 0 \\
\hline
\mathcal{V}_{n+1,1} & \cdots & \mathcal{V}_{n+1,n}  & 1 & \cdots & 0 \\
\vdots & \ddots & \vdots  & \vdots & \ddots & \vdots \\
\mathcal{V}_{m,1} & \cdots & \mathcal{V}_{m,n}  & 0 & \cdots & 1 \\
\end{array}
\right).
\end{equation}
\end{definition}

\begin{theorem}
    \label{thm:voting_matrix_convergence}
    Let \( \mathcal{V}\) be a voting matrix. Then, the sequence of its powers \( \mathcal{V}^x \) converges as \( x \to \infty \). Specifically, there exists a matrix \( \mathcal{W} \) such that:
    \begin{equation}
        \lim_{x \to \infty} \mathcal{V}^x = \mathcal{W}.
        \label{eq:vots_lim}
    \end{equation}
\end{theorem}

The proof of this theorem follows from the properties of \(\mathcal{V}\) as a probability transition matrix \cite{grinstead1997introduction}.\footnote{Note that the convergence of this matrix ($\mathcal{W}$) does not mean there will be a social choice, or social welfare.} 

\begin{corollary}[Optimized Computation of the Limit Matrix]
\label{thm:eff_comp_limit_matrix}
    Let \( \mathcal{V} \) be the voting matrix defined in Theorem \ref{thm:voting_matrix_convergence}. The calculation of the limit matrix \( \mathcal{W} \) can be optimized by considering a specific subsequence of matrix powers. Instead of evaluating the entire sequence \( \{\mathcal{V}^x\}_{x=1}^\infty \), the same limit \( \mathcal{W} \) can be obtained by computing the subsequence \( \{\mathcal{V}^{2^x}\}_{x=1}^\infty \), where \( \{2^x\}_{x=1}^\infty \) denotes the sequence of powers of two. Formally, the limit matrix \( \mathcal{W} \) is given by:
    \begin{equation}
        \mathcal{W} = \lim_{x \to \infty} \mathcal{V}^{2^x}.
    \end{equation}
    
    This approach exploits the exponential growth of \( 2^x \), reducing the number of matrix multiplications required to compute \( \mathcal{V}^{2^n} \) and thus reducing the overall computational complexity. Furthermore, this method preserves the numerical stability of the computation by mitigating the accumulation of rounding errors typically associated with iterative matrix multiplications. As a result, the subsequence-based approach provides a computationally efficient and robust mechanism for determining \( \mathcal{W} \), particularly in large-scale systems.
\end{corollary}

\subsection{Step 3: Generalization with Intermediaries}\label{sec:gPPV}

In formulating the matrix in \eqref{eq:voting_matrix_expanded}, we considered two domains, the voters and the proposals. We can generalize this to include a set of nodes or intermediate domains, such as political parties (a set of voters) and policy packages (a set of policies). These intermediaries can also be a target for partial votes.

\begin{itemize}
    \item Let \( X = \{x_1, x_2, \ldots, x_d\} \) represents the set of voters, where each \( x_i \) is an individual voter.
    \item Let \( Y = \{y_1, y_2, \ldots, y_m\} \) denotes the set of intermediaries, with each \( y_j \) representing an intermediary.
    \item Let \( Z = \{z_1, z_2, \ldots, z_p\} \) is the set of proposals, where each \( z_k \) is a single proposal.
\end{itemize}
The voting matrix \( \mathcal{V} \) in this system can be represented as:

\begin{equation} \label{eq:voting_matrix}
\mathcal{V} = \left(
\begin{array}{c|c|c}
  \mathcal{V}_{XX} & \mathcal{V}_{XY} & \mathcal{V}_{XZ} \\ \hline
  \mathcal{V}_{YX} & \mathcal{V}_{YY} & \mathcal{V}_{YZ} \\ \hline
  \mathcal{V}_{ZX} & \mathcal{V}_{ZY} & \mathcal{V}_{ZZ}
\end{array}
\right)
\end{equation}

putting all the discussion above $\mathcal{V}$ will look like this:

\begin{equation}
\mathcal{V} = \left(
\begin{array}{ccc|ccc|ccc}
  \mathcal{V}_{x_1x_1} & \cdots & \mathcal{V}_{x_dx_1} & 
  \mathcal{V}_{y_1x_1} & \cdots & \mathcal{V}_{y_mx_1} & 
  0 & \cdots & 0 \\
  \vdots & \ddots & \vdots & \vdots & \ddots & \vdots & \vdots & \ddots & \vdots \\
  \mathcal{V}_{x_1x_d} & \cdots & \mathcal{V}_{x_dx_d} & \mathcal{V}_{y_1x_d} & \cdots & \mathcal{V}_{y_mx_d} & 0 & \cdots & 0 \\ \hline

  \mathcal{V}_{x_1y_1} & \cdots & \mathcal{V}_{x_dy_1} & 
  \mathcal{V}_{y_1y_1} & \cdots & \mathcal{V}_{y_my_1} & 
  0 & \cdots & 0 \\
  \vdots & \ddots & \vdots & \vdots & \ddots & \vdots & \vdots & \ddots & \vdots \\
  \mathcal{V}_{x_1y_m} & \cdots & \mathcal{V}_{x_dy_m} & \mathcal{V}_{y_1y_m} & \cdots & \mathcal{V}_{y_my_m} & 0 & \cdots & 0 \\ \hline

  \mathcal{V}_{x_1z_1} & \cdots & \mathcal{V}_{x_dz_1} & 
  \mathcal{V}_{y_1z_1} & \cdots & \mathcal{V}_{y_mz_1} & 
  1 & \cdots & 0 \\
  \vdots & \ddots & \vdots & \vdots & \ddots & \vdots & \vdots & \ddots & \vdots \\
  \mathcal{V}_{x_1z_p} & \cdots & \mathcal{V}_{x_dz_p} & \mathcal{V}_{y_1z_p} & \cdots & \mathcal{V}_{y_mz_p} & 0 & \cdots & 1
\end{array}
\right)
\label{eq:voting_matrix_corrected}
\end{equation}

This generalization accommodates a wider range of voting systems, including hierarchical structures such as political parties or policy categories, while maintaining the coherence of the original framework.

\begin{figure}[ht]
  \centering
  \includegraphics[width=0.9\linewidth]{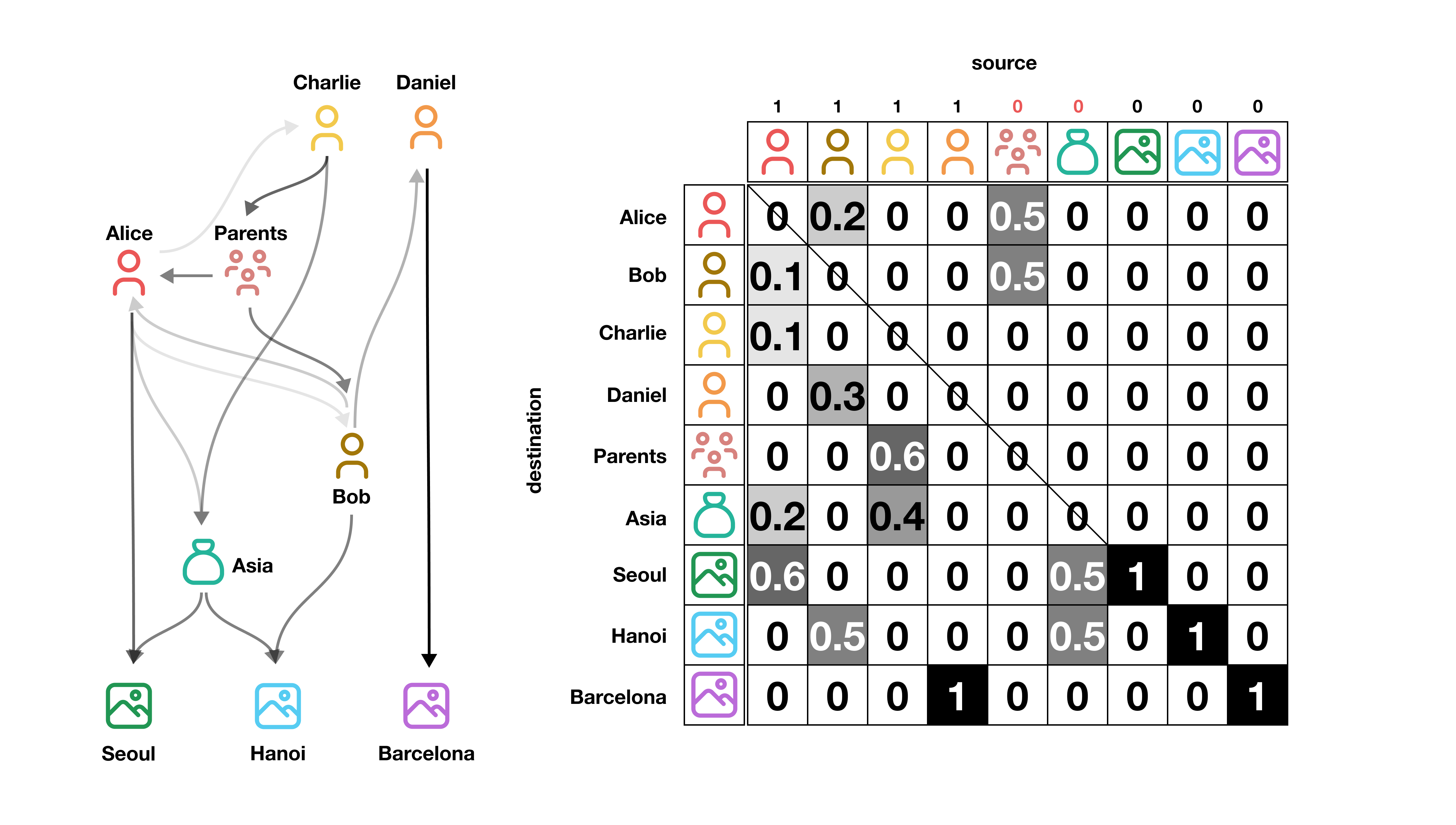}
  \caption{An example of PPV in action. The left shows the information as a network graph structure. Edges indicate partial trust between Nodes which are individuals like Alice, Bob, Charlie, and the parents group, family trip locations: Seoul, Hanoi, and Barcelona, and the category 'Asia'. 
The accompanying matrix on the right explains the identical information with real values ranging from 0 to 1. Notably, Alice (red) has given a higher preference (0.6) to visit Seoul, while Bob shows a preference for Hanoi, but delegating 30\% and 20\% of voting power to Daniel and Alice respectively. It indicates varying preferences among individuals, note that the sum of the columns is $1.0.$}
  \label{fig:ppvs_diagram}
\end{figure}

Figure \ref{fig:ppvs_diagram} shows a small example of a PPV session and a matrix looks like with four participants, three policies, and two intermediaries. The right is the social network which holds information identical to the left adjacent matrix.

As the structure of the matrix is not modified, all the properties analyzed in \ref{def:PPV} are maintained for the new voting matrix \eqref{eq:voting_matrix_corrected}.

\subsection{Influence: Net Proxy Vote}

Aside from knowing the final stable state for the consensus, it will be useful to know how much influence that user, group, or category had on that result. 
To quantify the total cumulative influence of voter \(i\) over an infinite horizon (as \(t \to \infty\)), we define

\begin{equation}\label{eq:net_proxy_votes}
    p_i = \frac{\sum\limits^H_{j=1}\sum\limits^{\infty}_{t=0}\mathcal{W}_{ij}(t)}{\sum\limits^{\infty}_{t=0}\mathcal{W}_{ii}(t)} \quad \mathrm{for}\ \mathrm{all}\ 1\leq i\leq N
\end{equation}

In Equation~\eqref{eq:net_proxy_votes}, the numerator aggregates the total votes a voter or policy category or group \(i\) receives from all other voters and intermediaries over all periods, while the denominator, reflecting the self-delegation votes accumulated by \(i\) acts as a normalization factor that neutralizes the effect of cyclic delegations.
 
This information is useful because it identifies key decision-makers within the network, as high values of \(p_i\) naturally highlight those voters who attract significant delegations, designating them as central figures or concepts in the decision-making process. Moreover, the metric is widely applicable to various networked decision-making systems, including our liquid democracy frameworks and organizational voting structures, where understanding the flow and concentration of delegated power is essential. Ultimately, the insights gained from the Net Proxy Vote analysis \(p_i\) can inform the design of more equitable vote redistribution algorithms, thus mitigating the risk of excessive power concentration and fostering a more balanced system.

This net proxy vote $p_i$, as defined in Equation~\ref{eq:net_proxy_votes}, quantifies the influence of each voter.

\begin{lemma}[Parallelism of Linear Functionals under a Rank-One Update]
Let $V$ be an $n \times n$ matrix and let $f \in \mathbb{R}^{1 \times n}$ be a row vector with the corresponding column vector $g = f^\top \in \mathbb{R}^{n \times 1}$. Assume that $f\cdot g \neq 0$, and define a rank-one modification of $V$ by
\[
V' = V \left( I - \frac{g\cdot f}{f\cdot g} \right).
\]
Suppose that both $I - V$ and $I - V'$ are invertible. Define
\begin{equation}
\label{eq:w_w'}
w = f (I - V)^{-1}, \quad w' = f (I - V')^{-1}.
\end{equation}

Then the vectors $w$ and $w'$ are parallel, and we have
\begin{equation}
\label{eq:lemma_parallel}
w' = \frac{f\cdot g}{w\cdot g} w.
\end{equation}

\end{lemma}

This result captures how a rank-one projection affects the transformation of a linear function under matrix inversion. It arises, for example, in the analysis of quasi-stationary behavior in Markov chains, where $V$ and $V'$ may represent sub-stochastic kernels over transient states.
 
\begin{proof}
Let $h$ be a column vector such that $f \cdot h = 0$.  
Then, using the definition of $V'$, we compute:
\[
V' h = V \left( h - \frac{g \cancel{f\cdot h}}{f \cdot g} \right) = Vh
\]

Recall (\ref{eq:w_w'}). So we have:

\begin{equation}
f \cdot h = w (I - V) h = w' (I - V') h = w' (I - V) h.
\end{equation}

Therefore, both $w$ and $w'$ are orthogonal to all vectors of the form $(I - V) h$,  
for $h$ in the null space of $f$ (an $(n-1)$ dimensional space).

Since $(I - V)$ is invertible, the image of this null space under $(I - V)$  
is also $(n-1)$-dimensional. Thus, both $w$ and $w'$ are in its 1-dimensional orthogonal complement, which means that $w$ and $w'$ are parallel.

Now, to compute the scaling factor, observe:
\[
V' g = V \left( g - \frac{g f \cdot g}{f \cdot g} \right) = V (g - g) = 0,
\]
so:
\begin{equation}
(I - V') g = g \quad \Rightarrow \quad w' \cdot g = f \cdot g.
\end{equation}

Since $w' = c w$ for some scalar $c$, we have:
\[
w' \cdot g = c (w \cdot g) \quad \Rightarrow \quad c = \frac{f \cdot g}{w \cdot g},
\]
and thus:
\[
w' = \frac{f \cdot g}{w \cdot g} \cdot w.
\]
\end{proof}

\begin{proof}[Net Proxy Vote]
In this section, we assume \( 1 \leq i \leq N \). Let \( f = e_i^\top \in \mathbb{R}^{1 \times N} \) be a row vector with a single 1 in position \( i \), and let \( g = e_i \in \mathbb{R}^{N \times 1} \) be the corresponding column vector. That is,
\begin{equation}
    f = (0, \dots, 0, 1, 0, \dots, 0), \quad g = \begin{pmatrix} 0 \\ \vdots \\ 0 \\ 1 \\ 0 \\ \vdots \\ 0 \end{pmatrix},
\end{equation}

where the 1 appears in the \( i \)-th entry. For the voting matrix \( \mathcal{V} \), the vector \( f \) can be used to extract the \( i \)-th row (i.e., the votes delegated \textit{to} agent \( i \)), while \( g \) selects the \( i \)-th column (i.e., the votes \textit{from} agent \( i \)).

We define a new matrix \( \mathcal{V}' \) that cancels the influence of incoming votes to node \( i \), effectively treating it as a terminal state. This is done via a rank-one update:

\begin{equation}
\mathcal{V}' = \mathcal{V} \left(I - \frac{g f}{f g} \right),
\label{eq:netproxy_matrix}
\end{equation}

which removes all incoming paths to node \( i \). Specifically, \( \mathcal{V}' \) is identical to \( \mathcal{V} \), except that the \( i \)-th column is replaced by zeros:

\begin{equation}
\mathcal{V}' =
\begin{pmatrix}
\mathcal{V}_{11} & \cdots & 0 & \cdots & \mathcal{V}_{1M} \\
\vdots & \ddots & \vdots &        & \vdots \\
\mathcal{V}_{M1} & \cdots & 0 & \cdots & \mathcal{V}_{MM}
\end{pmatrix}.
\label{eq:netproxy_columnzero}
\end{equation}

In this formulation, \( \mathcal{V}' \) behaves as a modified transition matrix in which node \( i \) receives no further incoming votes. In contrast, its votes are still distributed according to the original matrix \( \mathcal{V} \). In other words, agent \( i \) behaves as a \textit{proposal node}, or absorbing state in the Markov chain, accumulating all votes it receives without further delegation.

This yields the \textit{Net Proxy Vote} for agent \( i \), which corresponds to the total influence it holds when it votes directly, and cannot be used as a proxy by others.

Give row vectors $w$ and $w'$ the following:
\begin{align}
     w&= f+f\cdot \mathcal{V}+f\cdot \mathcal{V}^2+\cdots \\
    &= f\cdot (I-\mathcal{V})^{-1}\\
    w'&= f+f\cdot \mathcal{V}'+f\cdot (\mathcal{V}')^2+\cdots \\
    &= f\cdot (I-\mathcal{V}')^{-1}
\end{align}

$w$'s $j$ th element $w_j$ is by definition,
\begin{equation}
    w_j = \sum^{\infty}_{t=0}\mathcal{W}_{ij}(t) 
\end{equation}

From (\ref{eq:lemma_parallel}), $w'=\frac{f\cdot g}{w\cdot g}w=\frac{w}{w_i}$, so

\begin{equation}
    w'_j = \frac{\sum\limits^{\infty}_{t=0}\mathcal{W}_{ij}(t)}{\sum\limits^{\infty}_{t=0}\mathcal{W}_{ii}(t)}
\end{equation}

As stated above, the sum of $w'$ is the $i$th nodes Net Proxy Vote $p_i$, which is

\begin{equation}
    p_i = \frac{\sum\limits^H_{j=1}\sum\limits^{\infty}_{t=0}\mathcal{W}_{ij}(t)}{\sum\limits^{\infty}_{t=0}\mathcal{W}_{ii}(t)} \quad \mathrm{for}\ \mathrm{all}\ 1\leq i\leq N
\end{equation}

\end{proof}

\section{Utility Function}\label{sec:utility}
The goal of this section is twofold: first, to define a utility function that models how voters delegate their votes according to the dimensions described in Section \ref{sec:ppv}; and second, to use this utility function to establish a system that iterates over multiple discrete periods and tracks both the influence of each voter and the patterns of vote delegation.

\subsection{Utility Function Definition}
From our definitions in~\ref{def:LD} and~\ref{def:PPV}, recall the set of voters $d \in N$ (with $|N| = n$), and let policies be represented as $p \in P$. The set of alternatives is defined as the union of voters and policies: $A = N \cup P$ (where $|A| = m$). We denote voter $i$ as the delegatee and voter $j$ as the delegator (where $j$ assigns their vote to $i$). The vote assigned from $j$ to $i$ is represented by $D_{i,j}$. Additionally, we define the policy preferences of voter $j$ as a function:
\begin{equation}
    F_j \colon P \to \mathbb{R} \quad \quad \forall j \in N,
    \label{eq:Fj}
\end{equation}
where $F_j(p)$ quantifies $j$'s preference for policy $p \in P$.

\subsection{Simplified Utility Function}
The voter's utility function combines four components: (1) \textit{policy-aligned utility} $U_{\text{pref},i(p)} \in \mathbb{R}$, representing satisfaction when policy $p \in P$ matches their preferences $F_i \colon P \to \mathbb{R}$; (2) \textit{delegation gains} from expertise $e_{i,j} \geq 0$ and preference similarity $F_{i,j} = -\theta \|F_i - F_j\|$ ($\theta > 0$), where $\|\cdot\|$ measures preference distance; (3) \textit{control-loss cost} $C_j = \gamma \cdot \ell(j)$ ($\gamma > 0$), penalizing long delegation chains (with $\ell(j)$ as chain length); and (4) \textit{temporal dynamics} across periods $k = 1,\dots, K$. This models the trade-off between policy congruence, trust, and democratic participation.

\subsection{Constant Elasticity of Substitution Function for Vote Delegation}
The approach to modeling utility in PPV is to adopt a Constant Elasticity of Substitution (CES) function \cite{mcfadden1963constant}, commonly used in microeconomics. To tailor the CES function for vote delegation, consider that the delegator voter \(j\) "spends" their vote on delegatee voters \(i \in \{1, \dots, n\}\) (including delegating to themselves for direct voting). Thus, we define the voter delegation CES function as:
\begin{small}
\begin{equation}
\max_{y_1, y_2, \dots, y_n} U_j(y_{1,j}, y_{2,j}, \dots, y_{n,j}) = \left( \sum_{i=1}^{n} \alpha_{i,j}\, y_{i,j}^{\rho} \right)^{\frac{1}{\rho}},
\label{eq:ces_deleg}
\end{equation}
where we set
\begin{equation}
y_{i,j} = D_{i,j},
\label{eq:yD_relation}
\end{equation}
\end{small}
meaning that the vote assigned by \(j\) to \(i\) (weighted by the degree of agreement \(F_{i,j}\)) is represented by \(D_{i,j}\). The preference weight is defined as
\begin{equation}
\alpha_{i,j} = g(F_{i,j}, e_{i,j}),
\label{eq:alpha_def}
\end{equation}
and is subject to the constraints:
\begin{equation}
\sum_{i=1}^{n} \alpha_{i,j} = 1.
\label{eq:alpha_sum}
\end{equation}
In addition, we impose the delegation budget constraint:
\begin{equation}
\sum_{i=j+1}^{n} D_{i,j} \leq 1 + \sum_{j=1}^{n} D_{j,i}.
\label{eq:budget_deleg}
\end{equation}
This ensures that the total votes delegated by \(j\) do not exceed their inherent vote (1) plus any votes received from other voters.

\subsection{Incorporating Influence into the Utility Function}
To capture the benefit of maintaining influence, we introduce an additional term into the utility function. The influence that voter \(j\) receives in a period is measured as
\begin{equation}
\sum_{i=1}^{d} D_{j,i}.
\label{eq:influence_measure}
\end{equation}
To simulate diminishing marginal returns and ensure that the influence remains non-negative, we apply a logarithmic function. The modified utility function thus becomes:
\begin{small}
\begin{align}
\max_{y_1, y_2, \dots, y_n} U_j(y_{1,j}, \dots, y_{n,j}) 
&= \beta \left( \sum_{i=1}^{n} \alpha_{i,j} y_{i,j}^{\rho} \right)^{\frac{1}{\rho}} \nonumber \\
&\quad + \gamma \log \left(1 + \sum_{i=1}^{d} D_{j,i} \right)
\label{eq:ces}
\end{align}
\end{small}

\begin{equation}
\beta + \gamma = 1.
\label{eq:beta_gamma}
\end{equation}

\subsection{Dynamic Model in Discrete Periods}
PPV could be used for multiple rounds to record changes in voting behavior throughout the session. We distinguish the iteration for Markov chain convergence and multiple rounds here. Extending the model to a dynamic setting over discrete periods, the utility function for voter \(j\) in period \(k\) is given by:
\begin{small}
\begin{align}
\max_{\{y_{i,j}\}_{i=1}^n} \; U_j^k(y_{1,j}, \dots, y_{n,j}) 
&= \left( \sum_{i=1}^{n} \alpha_{i,j}^k \left( y_{i,j}^k \right)^{\rho} \right)^{\frac{1}{\rho}} \nonumber \\
&\quad + \gamma \log\left( 1 + \sum_{i=1}^{d} D_{j,i}^k \right),
\label{eq:ces_dynamic}
\end{align}
\end{small}

In Equation~\eqref{eq:ces_dynamic}, \(U_j^k\) denotes the utility of voter \(j\) in period \(k\), with
\begin{equation}
y_{i,j}^k = D_{i,j}^k,
\label{eq:yD_dynamic}
\end{equation}
and the weight coefficients are updated as:
\begin{equation}
\alpha_{i,j}^k = g\Bigl(y_{i,j}^{k-1}, F_{i,j}^{k-1}, F_{i,j}^k, e_{i,j}^k\Bigr).
\label{eq:alpha_dynamic}
\end{equation}
Furthermore, the vote redistribution function for the next period is defined by:
\begin{equation}
D_{i,j}^k = \frac{\alpha_{i,j}^k\, \Bigl(F_{i,j}^k\Bigr)^{\rho}}{\sum_{l=1}^{n} \alpha_{l,j}^k\, \Bigl(F_{l,j}^k\Bigr)^{\rho}}, \quad \text{for all } j = 1, \dots, d.
\label{eq:deleg_redistribution}
\end{equation}

In period \(k\), after computing the dynamic utility function (see Equation~\eqref{eq:ces_dynamic}), each voter \(i\) receives a total delegated vote given by \(D_{i,j}^k\). To account for changes in policy preferences between rounds, voters select a policy \(p\) to form the outcome set \(A\). The utility derived from policy outcomes is redefined as:
\begin{equation}
U_{pref,j}^k = U_{pref,j}^k\Bigl(F_j^k, A^k\Bigr).
\label{eq:Upref_k}
\end{equation}
Moreover, it is assumed that these outcomes affect future policy preferences according to:
\begin{equation}
F_j^k = F_j^k\Bigl(A^{k-1}\Bigr).
\label{eq:Fj_update}
\end{equation}
(See Equation~\eqref{eq:Fj_update} for the update rule of the policy preferences.)

\subsection{Total Utility}
The total utility obtained by voter \(j\) in period \(k\) is expressed as the sum of the maximized delegation utility and the utility derived from policy outcomes:
\begin{equation}
\sum_{j=1}^{n} \Bigl[\max_{y_1, y_2, \dots, y_n} U_j^k(y_{1,j}, y_{2,j}, \dots, y_{n,j}) + U_{pref,j}^k\Bigr].
\label{eq:total_utility_k}
\end{equation}
Over a total of \(z\) periods, the cumulative utility for all voters is given by:
\begin{equation}
\sum_{k=1}^{z} \sum_{j=1}^{n} \max_{y_1, y_2, \dots, y_n} U_j^k(y_{1,j}, y_{2,j}, \dots, y_{n,j}).
\label{eq:total_utility_z}
\end{equation}

\section{Case Study: PPV in action}\label{sec:validation}

Based on the formalization so far, a resource allocation experiment using the above-mentioned model was conducted using the Cambridge Participatory Budgeting(CPB) for the year 2023 as the topic. CPB had 20 proposals from which residents can choose. Each proposal had a title, description, cost, location, and category. The 20 proposals were evenly distributed into four categories. We adopted CPB as a topic and used PPV to evaluate which plan should be executed with 69 participants. The participants were divided into 10 groups.

\begin{figure}[ht]
  \centering
  \includegraphics[width=0.5\linewidth]{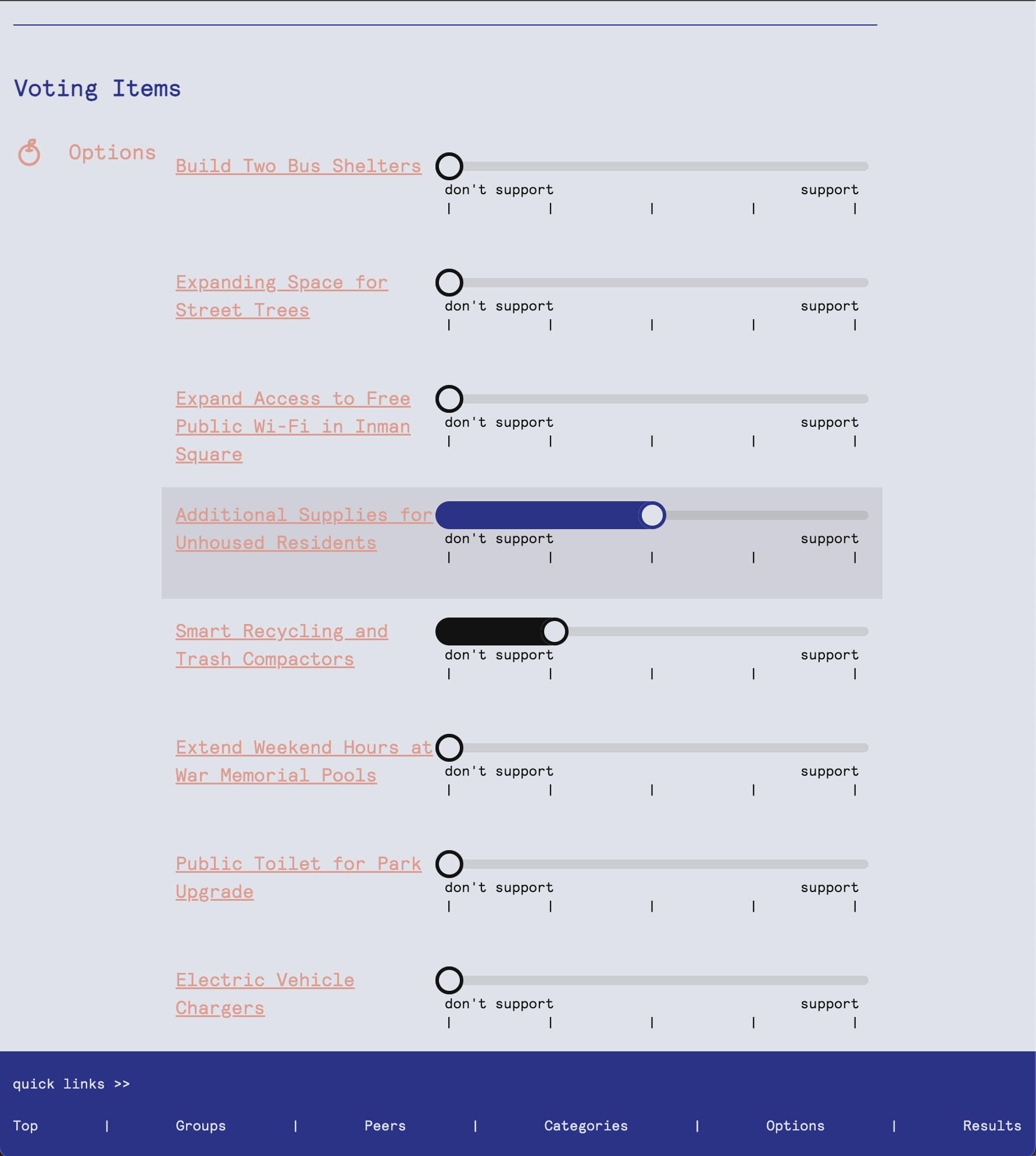}
  \caption{The PPV webapp interface. Participants individually rated using a slider interface.}
  \label{fig:webapp}
\end{figure}

In the context of this paper, tables, and categories can be considered intermediaries. 

after a presentation on the evaluation method and the topic, participants were asked to discuss within their table groups for 5 minutes followed by a 2-minute presentation to the whole room. The participants then interacted with a web application Fig.\ref{fig:webapp} which they could freely rate each proposal, category, peer, and table. Each evaluation was performed individually.

\subsection{Results}
Tables \ref{tab:influence} and \ref{tab:consensus} show the results of the calculation after the workshop. The ranking (Table \ref{tab:consensus}) shows which policy gained how many fractional votes. 

\begin{table}[!h]
\centering
\caption{Influence Score among Participants, Groups, and Categories. List showing the top 15 nodes of influences. The result shows that the Environment category had the most fractional delegation passed through. As for the type, Categories have dominated this metric, which is useful information for further deliberation. The minimal value of this metric will be 1.0, coming from the fact that there was no partial delegation to that node.}
\label{tab:influence}
\begin{tabular}{cllcl}
\toprule
\textbf{Rank} & \textbf{Entity Name} & \textbf{Score} & \textbf{Type} \\
\midrule
1  & Environment & 4.588 & Category \\
2  & Transportation, Streets and Sidewalks & 4.370 & Category \\
3  & Community Resources & 4.360 & Category \\
4  & Facilities, Parks, and Recreation & 3.430 & Category \\
5  & Team 8 & 3.103 & Team \\
6  & Team 10 & 3.005 & Team \\
7  & M. I & 2.569 & User \\
8  & I. A & 2.362 & User \\
9  & Team 3 & 2.190 & Team \\
10 & Team 9 & 2.139 & Team \\
11 & Team 1 & 2.097 & Team \\
12 & A. S & 1.933 & User \\
13 & Team 6 & 1.876 & Team \\
14 & Team 5 & 1.854 & Team \\
15 & Team 2 & 1.588 & Team \\
\bottomrule
\end{tabular}
\end{table}

The sum of all policy scores is $69$ which the one-person one-vote policy still holds. The influence score (Table \ref{tab:influence}) is the 15 top influential nodes among peers, categories, and groups. 

\begin{table*}
  \caption{Favored proposals using Propagational Proxy voting. Note: the sum of all votes equals 69.0, exemplifying that the one-person-one-vote principle still holds.}
  \label{tab:consensus}
  \begin{tabular}{l|c}
    \toprule
    Proposals & Votes \\
    \midrule
    Smart Recycling and Trash Compactors                                    & 4.60 \\  
    Free Menstrual Care and Infant Hygiene Products                         & 4.50 \\     
    Improve Safety for Pedestrians                                          & 4.18 \\ 
    Public Toilet for Park Upgrade                                          & 4.12 \\ 
    Electric Vehicle Chargers                                               & 3.99 \\ 
    Floating Bus Stops: Improved Safety for Cyclists and Bus Riders         & 3.77 \\ 
    Additional Supplies for Unhoused Residents                              & 3.65 \\ 
    Shaded Seats on Hot Streets                                             & 3.64 \\ 
    Expand Access to Free Public Wi-Fi in Inman Square                      & 3.57 \\ 
    Build Two Bus Shelters                                                  & 3.49 \\ 
    Expanding Space for Street Trees                                        & 3.42 \\ 
    Smart Traps for Rat Reduction                                           & 3.15 \\ 
    Teen Oasis at Donnelly Field                                            & 3.00 \\ 
    Add and Maintain Bike Repair Stations                                   & 2.94 \\ 
    Tree and Community Garden Engagement Coordinator                        & 2.93 \\ 
    Expand Access to Veterinary Care for Seniors and Low-Income Communities & 2.90 \\ 
    Let’s Help Cambridge Residents Buy Bicycles                             & 2.89 \\ 
    The Cambridge Women’s Heritage Project                                  & 2.88 \\ 
    Extend Weekend Hours at War Memorial Pools                              & 2.32 \\ 
    Sand Volleyball Court                                                   & 2.30 \\ 
    \bottomrule
  \end{tabular}%
\end{table*}

We can see that categories gained the most delegation compared to other types such as Teams and Users. This information is useful for orienting the discussion; in this case, the organizer could ask for additional comments on the environmental category to further understand what led to this outcome.

\section{Discussion}\label{sec:discussion}

\subsection{LD as a special case and PPV's extendability}
When defining the model, we started to model LD using Markov chains. We then used the same framework to extend it to PPV (by relaxing \eqref{eq:LD_binary_constraint} to \eqref{eq:PPV_fraction}). With this relaxation, PPV is further extensible on how we obtain input from participants. One could think there will be a ranked order choice that will be transferred as Borda count\cite{emerson2013original} method, or have Quadratic Voting \cite{lalley2016quadratic} to modify how the fractional votes will be incorporated into the voting matrix of PPV. That being said, the fractional vote allocation for intermediaries will remain a source of discussion. The case study \ref{sec:validation} used an equal vote distribution for intermediaries, yet there could be any number of strategies to allocate internal vote allocation between groups and categories. Real-world examples can be seen in how political parties distribute their gained votes to the members of that party.

\subsection{Comparing to liquid democracy}
As we discussed in the model, PPV is an extension of LD, which we can simulate the LD result from the PPV input. Table \ref{tab:vsld} is the comparison between the two approaches. In aggregated view, the standard deviations for PPV and LS are $0.6417$ and $4.2007$, respectively, which has a significant difference. Although we cannot generalize this, we have observed that PPV's result has a less extreme result, which may come from the expressivity of PPV compared to LD.

\begin{table*}[!ht]
\scriptsize
\centering
\caption{Comparison of results by PPV versus the Liquid Democracy (LD) method. PPV integrates delegation chains and influence propagation into its scoring, yielding rankings and scores that differ markedly from those obtained under LD. The columns “PPV Ranking” and “PPV Score” present outcomes from the PPV method, while “LD Ranking” and “LD Score” show the respective values under LD. The $\Delta$ columns highlight the impact of PPV’s delegation and influence mechanisms compared to the direct vote aggregation in LD. The LD result was calculated using the same input from PPV by (1) removing intermediary nodes, and (2) assigning a full vote to the top-ranked item per participant. In the case of ties, votes were split equally. The standard deviations of the votes were $0.6417$ for PPV and $4.2007$ for LD.}
\label{tab:vsld}
\resizebox{\textwidth}{!}{
\begin{tabular}{l|c|l|c|c|c}
\toprule
\textbf{PPV Proposal} & \textbf{PPV Score} & \textbf{LD Proposal} & \textbf{LD Score} & $\Delta$ \textbf{Rank} & $\Delta$ \textbf{Score} \\
\midrule
Smart Recycling and Trash Compactors & 4.60 & Improve Safety for Pedestrians & 15.89 & +2 & +11.72 \\
Free Menstrual Care and Infant Hygiene Products & 4.50 & Free Menstrual Care and Infant Hygiene Products & 10.66 & +0 & +6.16 \\
Improve Safety for Pedestrians & 4.18 & Public Toilet for Park Upgrade & 9.69 & +1 & +5.57 \\
Public Toilet for Park Upgrade & 4.12 & Smart Recycling and Trash Compactors & 7.49 & -3 & +2.89 \\
Electric Vehicle Chargers & 3.99 & Electric Vehicle Chargers & 4.79 & +0 & +0.80 \\
Floating Bus Stops & 3.77 & Shaded Seats on Hot Streets & 4.33 & +2 & +0.69 \\
Additional Supplies for Unhoused Residents & 3.65 & Additional Supplies for Unhoused Residents & 3.05 & +0 & -0.60 \\
Shaded Seats on Hot Streets & 3.64 & Teen Oasis at Donnelly Field & 2.46 & +5 & -0.53 \\
Expand Access to Free Public Wi-Fi in Inman Square & 3.57 & Build Two Bus Shelters & 2.34 & +1 & -1.15 \\
Build Two Bus Shelters & 3.49 & Expand Access to Free Public Wi-Fi in Inman Square & 2.13 & -1 & -1.44 \\
Expanding Space for Street Trees & 3.42 & Floating Bus Stops & 2.02 & -5 & -1.74 \\
Smart Traps for Rat Reduction & 3.15 & The Cambridge Women’s Heritage Project & 1.25 & +6 & -1.63 \\
Teen Oasis at Donnelly Field & 3.00 & Sand Volleyball Court & 0.98 & +7 & -1.32 \\
Add and Maintain Bike Repair Stations & 2.94 & Veterinary Care for Seniors & 0.64 & +2 & -2.25 \\
Tree and Community Garden Engagement Coordinator & 2.93 & Smart Traps for Rat Reduction & 0.50 & -3 & -2.65 \\
Veterinary Care for Seniors & 2.90 & Street Trees Expansion & 0.33 & -5 & -3.08 \\
Let’s Help Cambridge Residents Buy Bicycles & 2.89 & Pool Hours Extension & 0.25 & +2 & -2.07 \\
The Cambridge Women’s Heritage Project & 2.88 & Community Garden Coordinator & 0.17 & -3 & -2.76 \\
Pool Hours Extension & 2.32 & Bike Repair Stations & 0.00 & -5 & -2.94 \\
Sand Volleyball Court & 2.30 & Buy Bicycles Program & 0.00 & -3 & -2.89 \\
\bottomrule
\end{tabular}
}
\end{table*}

\subsection{Computational Load and Speed-Up for scale}\label{sec:computational}

It is important to note that while the PPV model offers a nuanced framework to capture the intricacies of vote delegation and influence propagation, it incurs a high computational cost. The iterative calculation of \(\mathcal{V}^x\) required to obtain the limit matrix $\mathcal{W}$ - and ultimately the Net Proxy Vote \(p_i\) - can be demanding in large-scale systems and could be addressed in two ways. For two-dimensional matrices, advances in computational algorithms have led to more efficient methods than the naive $O^3$. For example, the \textit{Strassen algorithm} \cite{strassen1969gaussian} reduces the complexity to approximately \(O(n^{2.807})\); more recent developments have even reduced the bound to approximately \(O(n^{2.3728596})\) \cite{alman2020refined}.

On the other hand, Singular Value Decomposition \textit{SVD} is widely used for data processing and signal analysis. However, SVD has a complexity close to \(O(n^3)\), and the subsequent exponentiation of the decomposed matrices adds further computational overhead. This makes the SVD-based approach less attractive for our scenario.

Although advanced techniques (e.g., Strassen and its successors) and hardware innovations (GPU) improve the efficiency of matrix multiplication, the overall computational load remains a crucial limitation. Addressing these challenges is essential to ensure that the rich theoretical framework of PPV can be scaled to real-world applications. 

\section{Conclusion} \label{sec:conclusion}
This paper introduced Propagational Proxy Voting (PPV), an extension of Liquid Democracy that allows for fractional delegation of votes between proposals, individuals, and intermediary sets. By modeling the voting process using absorbing Markov chains, the PPV framework captures complex hierarchical structures and nuanced voter preferences that are often overlooked in traditional collective decision making. Our model demonstrated that this fractional vote propagation converges to a stable distribution, and we introduced a net influence metric to quantify the impact of voters beyond direct ballots. We also modeled the utility using the CES function, and how this trends in multiple rounds.

Our empirical study, conducted with $69$participants in Cambridge's participatory budgeting context, validated the applicability of the model and revealed how intermediary nodes, such as categories or groups, play a significant role in collective decision outcomes. The results show promise in integrating PPV into participatory platforms in the real world where diversity of opinion, knowledge asymmetry, and deliberative dynamics are critical.

Future research will focus on scaling the computational implementation of PPV to larger populations while exploring the behavioral impacts of diverse UI designs for fractional voting. Alongside the UI, we also plan to examine multi-round utility trends. Integrating PPV into live civic platforms holds promise for deepening democratic engagement in increasingly complex decision environments. In parallel, we plan to extend our analysis by incorporating advanced sensitivity and perturbation techniques, which will allow us to rigorously evaluate how small variations in delegation strategies impact the final influence distribution. Future research will also focus on user interfaces and examine which interface reflects the model while being approachable. Although these ideas are not fully developed here, their incorporation marks a promising avenue for both theoretical advancement and practical applications.

\section*{Acknowledgment}

The authors acknowledge that this research was advanced with generous support from the City Science Laboratory in Andorra and the MIT City Science network, where we have also conducted similar workshops with early versions of the application. Andrew Grace, Chris Osgood, and Nigel Jacob have provided valuable insight while developing this model on how this could contribute in real-world settings. The input not only helped the development, but will set a direction for future application domains. We also thank Allen Song for his support in providing feedback on the implementation of the model. The insights he gave towards the original implementation fortified our understanding. The authors give special thanks to the students who participated in the workshop.

\bibliography{ppvs}
\end{document}